\documentstyle[referee,psfig]{l-aa}

\input psfig

\begin{document}

   \thesaurus{11.07.1; 
              12.07.1} 

   \title{Magnification cross-sections of gravitational lensing 
	by galaxies in general FLRW cosmologies}

  \author{Zong-Hong Zhu\inst{1,2} }

  \offprints{Z.-H. Zhu\inst{1}}

   \institute{Institute of Theoretical Physics, 
		Chinese Academy of Sciences, Beijing 100080, China
	\and	Beijing Astrophysics Center (BAC)\thanks{BAC is jointly 
			sponsored by The Chinese Academy of Sciences and 
			Peking University},
		Beijing, 100871, China}

   \date{Received 00 00, 1998; accepted 00 00, 1998}

   \maketitle
   \markboth{Z.-H. Zhu: 
		Magnification cross-sections of gravitational lensing
		by galaxies in general FLRW cosmologies}{}

   \begin{abstract}

For a wide variety of cosmological models characterized by the
cosmic mass density $\Omega_M$ and the normalized cosmological 
constant $\Omega_{\Lambda}$,   we derive an analytic expression 
for the estimate of magnification cross-sections
by an ensemble of isothermal spheres as models of galactic mass
distributions. This provides a simple approach to demonstrate how the
lensing probability by galaxies depends on the cosmological
parameters. An immediate consequence is that,   
while a non-zero cosmological constant indeed leads to a significant 
increase of the lensing probability as it has been shown in the literature, 
only a small fraction of sky to $z\sim 3$ can be moderately ($\mu \sim 1.3$) 
lensed by galaxies  even in a $\Lambda$-dominated flat universe. 
Therefore, whether or not there is a nonzero cosmological 
constant,  it is unlikely that the overall quasar counts have been seriously
contaminated by the presence of galactic lenses.\\

  \keywords{gravitational lensing -- galaxies: general}

\end{abstract}

\section{Introduction}

It is well known that the gravitational lensing  by the foreground 
objects (e.g., galaxies) can alter the apparent brightness of background 
objects (e.g., quasars), which may contaminate our observations. 
Three decades ago, Barnothy and Barnothy (1968) proposed that
all the quasars were nothing but the gravitationally magnified
images of Seyfert galactic nuclei. Press and Gunn (1973) showed
that the probability of occurrence of gravitational lensing in an
$\Omega=1$ universe is nearly unity. For many years  there had been 
a lack of both convincing observational and theoretical supports
for these speculations. 
However, numerous and unprecedented deep galaxy surveys have
recently revealed a considerably large population of faint 
galaxies (Metcalfe et al. 1996; references therein). This
motivates one to readdress the question if
the observations of background objects are seriously affacted
by the gravitational lensing effect of foreground galaxies?
For this purpose, Zhu \& Wu (1997) have calculated the lensing
cross-sections of background quasars by the foreground galaxies, 
and concluded that, despite the fact
that there is a considerably high surface number density of faint galaxies 
the total lensing cross-sections by galaxies towards a distant source 
are still rather small, when only
a special cosmological model of $\Omega_M =1$ is considered.

Nevertheless, the optical depth (probability) of gravitational lensing 
depends sensitively on the cosmological models. 
It is worthy of examining whether the above claim is valid
under  general cosmological models. In this paper,
we extend our previous work to  a variety of 
cosmological models, which are characterized by the mass density parameter  
$\Omega_M$ and the normalized cosmological constant $\Omega_{\Lambda}$ 
(cf., Carroll, Press \& Turner 1992). This is well motivated because
cosmological models with nonzero cosmological constant have become quite
popular recently.
Many years ago, Gott, Park \& Lee (1989) have given the general expressions 
for the optical depth and mean image seperation in general
Friedman-Lema\^itre-Robertson-Walker (FLRW) cosmological models.
Yet, these expressions are complicated and thereby hard to use in practice.
One of the purposes of this paper is thus to simplify the 
formula. Furthermore, we would like to investigate how the cosmological
parameters affect the estimate of the lensing cross-section.

\section{Distance and volume measures in general FLRW cosmologies}

We assume a
homogeneous and isotropic universe described by a Robertson-Walker 
metric (Weinberg 1972):
\begin{equation}
ds^2 = -c^2dt^2 + R^2(t) \left[ \frac{dr^2}{1-kr^2} +
r^2(d \theta^2 + \sin^2\theta d\phi^2)\right].
\end{equation}
or in the form
\begin{equation}
ds^2 = -c^2dt^2 + R^2(t) \left[ d\chi^2 +
f(\chi)^2(d \theta^2 + \sin^2\theta d\phi^2)\right].
\end{equation}
where
\begin{equation}
\label{r}
r = f(\chi) = {\rm sinn}(\chi) \equiv \left\{ \begin{array}{ll}
        \sin\chi & k=+1\,,\\
        \chi & k=0\,,\\
        \sinh\chi & k=-1\,.
        \end{array}
        \right.
\end{equation}

The relation of the measurables to the unmeasurables is (Lightman et al. 1975;
Carroll, Press \& Turner 1992)
\begin{equation}
\label{distances}
        (1+z)=R(t_0)/R(t), \,\, D^A=R(t)r; \,\,
        D^M=R(t_0)r, \,\, D^L=R^2(t_0)r/R(t). 
\end{equation}
where $t_0$ is the present time of the universe, and $D^A$, $D^M$ and 
$D^L$ are the angular diameter distance, the
proper motion distance and the luminosity distance respectively.
Distances used in lensing theory are the angular diameter distances
(Schneider, Ehlers \& Falco 1992). From Eqs.~\ref{r} and \ref{distances}, 
one can derive an important relation
\begin{equation}
\label{ratio}
        \frac{{D^A}_{ds}}{{D^A}_s} = \frac{f(\chi_s - \chi_d)} {f(\chi_s)}
\,,
\end{equation}
where $D^A_{ds}$ and $D^A_s$ are the angular distances from the lens to the
source and from the observer to the source, respectively.

Using the Einstein field equation, it can be shown that the relation of the 
comoving distance $\chi$ to the redshift for the general FLRW cosmologies is
\begin{equation}
\begin{array}{ll}
\label{comoving}
        \chi & = \frac{c}{H_0 R(t_0)} \int_0^z
        \frac{dz}{\sqrt{(1+z)^2(1+\Omega_M z) - z(2+z) \Omega_{\Lambda}}} \\
{} & {}\\
        {}   & = \left\{ \begin{array}{ll}
                \frac{c}{H_0 R(t_0)} \int_0^z
                 \frac{dz}{\sqrt{\Omega_M (1+z)^3 - \Omega_M + 1}} ,
           & \Omega_M + \Omega_{\Lambda} = 1\,\, (k=0)\,,\\
                \left|\Omega_M + \Omega_{\Lambda} -1\right|^{1/2} \int_0^{z}
        \frac{dz}{\sqrt{(1+z)^2(1+\Omega_M z) - z(2+z)\Omega_{\Lambda}}},\,\,\,
        & \Omega_M + \Omega_{\Lambda} \neq 1 \,\, (k=\pm 1)\,.\\
                        \end{array}
                        \right.
\end{array}
\end{equation}

For our end, the comoving volume $dV$ is more convenient than the traditional 
physical volume. 
Within the shell $d\chi$ at $\chi$,  $dV$ reads (Gott, Park \& Lee 1989)
\begin{equation}
\label{volume}
        dV = 4\pi R^3(t_0) f^2(\chi)d\chi = \left(\frac{c}{H_0}\right)^3 
		4\pi \left(\frac{c}{H_0 R(t_0)}\right)^{-3} f^2(\chi)d\chi
\,.
\end{equation}

\section{Magnification cross-sections of gravitational lensing}

First of all, we consider the lensing cross-section (Turner, Ostriker \&
Gott 1984) due to a specific galaxy. 
Following Turner et al. (1984), we model the mass density profile of the total 
galaxy matter as the singular isothermal sphere (SIS), whose magnification 
for a point source is given by (Schneider, Ehlers \& Falco 1992; Wu 1996)
\begin{equation}
	\mu = \frac{\theta}{\theta - \theta_E}, \,\,\,\, 
			{\rm for}\,\,\, \theta > \theta_E
	\equiv 4\pi \left(\frac{\sigma}{c}\right)^2 \frac{D^A_{ds}}{D^A_s}
\,,
\end{equation}
where $\theta$ is the observed angular position of the source (image position),
$\theta_E$ is the angular radius of Einstein ring and 
$\sigma$ is the velocity dispersion of the lensing galaxy.
Note that we only include the contribution of the primary image 
because here we will not deal with
the statistics of multiple images. The dimensionless magnification
cross-section for a point source located at $z_s$ produced by a single SIS 
galaxy at $z_d$ is
\begin{equation}
        \hat{\sigma}(>\mu) = \frac{1}{{\left( \mu -1 \right)}^2}
                                \pi \theta^2_E
			= \frac{1}{\left( \mu -1 \right)^2}
                16{\pi}^3 \left( \frac{\sigma}{c}\right)^4
        \left[\frac{f(\chi_s - \chi_d)}{f(\chi_s)}\right]^2
\,,
\end{equation}
where the relation of Eq.~\ref{ratio} has been employed.

Now, let's consider the contributions of an ensemble of
galaxies having different luminosities and redshifts. 
The present-day galaxy luminosity function can be described by the Schechter 
function (Peebles 1993)
\begin{equation}
\phi_i(L)dL=\phi_i^*(L/L_i^*)^{-\alpha}\exp(-L/L_i^*)d(L/L_i^*),
\end{equation}
where $i$ indicates the morphological type of galaxies: $i$=(E, S0, S). 
The above expression can be converted into the velocity dispersion distribution
through the empirical formula between the luminosity and the central dispersion
of local galaxies $L/L_i^*=(\sigma/\sigma_i^*)^{g_i}$.
We keep the same parameters ($\phi_i^*, L^*_i, \alpha; \sigma_i^*, g_i$) 
as those adopted by Kochanek (1996) based on the surveys (Loveday
et al. 1992, Marzke et al. 1994), which yield 
$\sigma_i^*=(220,220,144)$ km/s and $g_i=(4,4,2.6)$ for $i=(E,S0,S)$ galaxies,
and the morphological composition $\{\gamma_i\}= 
(44, 56)$ for ($E+S0,S$).
For the spatial distribution of galaxies, we employ a general FLRW cosmological
model parametrized by $\Omega_M$ and $\Omega_{\Lambda}$, which has been
outlined in section 2. Finally, the total dimensionless
magnification cross-section by galaxies at redshifts ranging from 0 to 
$z_s$ for the distant sources like quasars at $z_s$ is
\begin{equation}
\label{total-cross}
    \hat{\Sigma}(z_s, >\mu) = 4\pi \frac{1}{{\left( \mu -1 \right)}^2}
                         \left( \sum_{i = E, S0, S} F_i \right)
                        T(z_s)
\,,
\end{equation}
The parameter $F_i$ represents the effectiveness of the $i$-th morphological 
type of galaxies in producing double images (Turner et al. 1984), which reads
\begin{equation}
\label{Fi-general}
F_i \equiv 16\pi^3 \left(\frac{c}{H_0}\right)^3
                        \langle n_{0i} \left(\frac{\sigma}{c}\right)^4 \rangle
        = 16\pi^3 \left(\frac{c}{H_0}\right)^3 \phi^* \gamma_i
		\left(\frac{b_i \sigma_i^*}{c}\right)^4 
		\int (L/L_*)^{\alpha + 4/g_i} \exp(L/L_*)dL/L_*
\,,
\end{equation}
where $b_i$ is the 
velocity bias between the velocity dispersion of stars and of dark matter 
particles. 
The above equation can be further written as 
\begin{equation}
\label{Fi-maximum}
F_i = 16\pi^3 \left(\frac{c}{H_0}\right)^3 \phi^* \gamma_i
	\left(\frac{b_i \sigma_i^*}{c}\right)^4 \Gamma(-\alpha+4/g_i+1), \,\,\,
	{\rm if} \,\,\, L \in (0,\infty)
\,,
\end{equation}
if the integral is performed from $0$ to $\infty$. 
In practice, the galaxy luminosities have the minimum and maximum limits,
and, therefore, Eq.~\ref{Fi-maximum} is the maximum estimate of $F_i$.
The $z_s$ dependent factor $T(z_s)$  is
\begin{equation}
        T(z_s) = \left(\frac{c}{H_0 R(t_0)}\right)^{-3}
		\int_0^{\chi_s} {\left[ \frac{f(\chi_s - \chi_d)}
                {f(\chi_s)} \right]}^2 f^2(\chi_d) d\chi_d
\,.
\end{equation}
For general FLRW cosmologies, an analytic expression is found:
\begin{equation}
\label{T-general}
T(z_s) = \left\{ \begin{array}{ll}
        \left(\frac{c}{H_0 R(t_0)}\right)^{-3} \frac{\chi^3_s}{30},
                & \Omega_M + \Omega_\Lambda =1\,,\\
        \left|\Omega_M + \Omega_{\Lambda} -1\right|^{-3/2}
        \left[\frac{1}{8}(1 + 3 \cot^2\chi_s) \chi_s -
        \frac{3}{8}\cot\chi_s\right], & \Omega_M + \Omega_\Lambda >1\,,\\
        \left|\Omega_M + \Omega_{\Lambda} -1\right|^{-3/2}
        \left[\frac{1}{8}(-1 + 3 \coth^2\chi_s) \chi_s -
       \frac{3}{8}\coth\chi_s\right], \,\,\, & \Omega_M + \Omega_\Lambda <1\,,\\
        \end{array}
        \right.
\end{equation}
where $\chi_s$ can be calculated through Eq.~\ref{comoving}.
For a flat universe ($\Omega_M + \Omega_\Lambda =1$), it reduces to
(Turner 1990)
\begin{equation}
\label{T-flat}
        T(z_s) = \frac{1}{30} {\left[\int_0^{z_s} \frac{dz}
                {\sqrt{\Omega_M (1+z)^3 - \Omega_M + 1}}
                \right]}^3, \,\,\, \Omega_M + \Omega_\Lambda =1
\,.
\end{equation}
If $\Omega = 1$ and $\Omega_\Lambda =0$, it reads (Turner et al. 
1984)
\begin{equation}
\label{T-Ede}
T(z_s) = \frac{4}{15}
            \frac{[(1+z_s)^{1/2}-1]^3}{(1+z_s)^{3/2}}, \,\,\,\,\,\, 
		\Omega_M = 1, \, \Omega_{\Lambda} =0.
\end{equation}

We should point out that the expression of Eq.~\ref{total-cross} is 
very useful. Dividing the expression by $4\pi$, 
one gets the fraction of the sky within redshift $z_s$ 
which is magnified by the factor greater than $\mu$:
\begin{equation}
\label{probability}
p(z_s, >\mu) = \frac{1}{(\mu - 1)^2} \, F \,\, T(z_s)
\,.
\end{equation}
We employ $F = \displaystyle\sum_{i = E, S0, S} F_i$ denoting the total 
effective parameter of all galaxies in producing multiple images. Further 
omitting the $\mu$-dependent term in Eq.~\ref{probability}, one obtains the 
conventional optical depth for multiple images.

In the above calculations, we have assumed that the comoving number
density of galaxies is constant. However, this may not hold true for the
realistic situation. The influence of galaxy evolution on the 
the lensing cross-section should also be taken into
account. Zhu and Wu (1997) have 
include this effect by using the galaxy merging model proposed by
Broadhurst et al. (1992), since 
the scenario of galaxy merging can account for both
the redshift distribution and the number counts of galaxies at
optical and near-infrared wavelengths (Broadhurst et al. 1992).
There are two effects arising from the galaxy merging: 
The first  is that there are more galaxies and hence more lenses in the past.
The second  is that galaxies are typically 
less massive in the past and hence less efficient as lenses.
As a result of two effects, the total magnification cross-section 
remains roughly unchanged (Zhu \& Wu 1997).

\section{Results and discussions}

Knowing the analytic expressions for the lensing cross-sections in general 
FLRW cosmologies, we can explore in detail the influences of cosmological
parameters  by numerical computation. 
We compute the probability $p(z_s, >\mu)$ so as to investigate whether 
the background objects (like quasars) counts are significantly 
contaminated. Eq.~\ref{probability} contains three factors, namely,
the $\mu$-dependent term, the galaxies term and the cosmological
term, associated with $p(z_s, >\mu)$. 
Here we concentrate on the cosmological term by adopting  
the maximum value $F\sim 0.028$ (Kochanek 1996)and a moderate
magnification of $\mu\sim 1.3$.

\begin{figure*}
\label{Figure 1}
\hbox{
\hspace*{0.6cm}
\psfig{figure=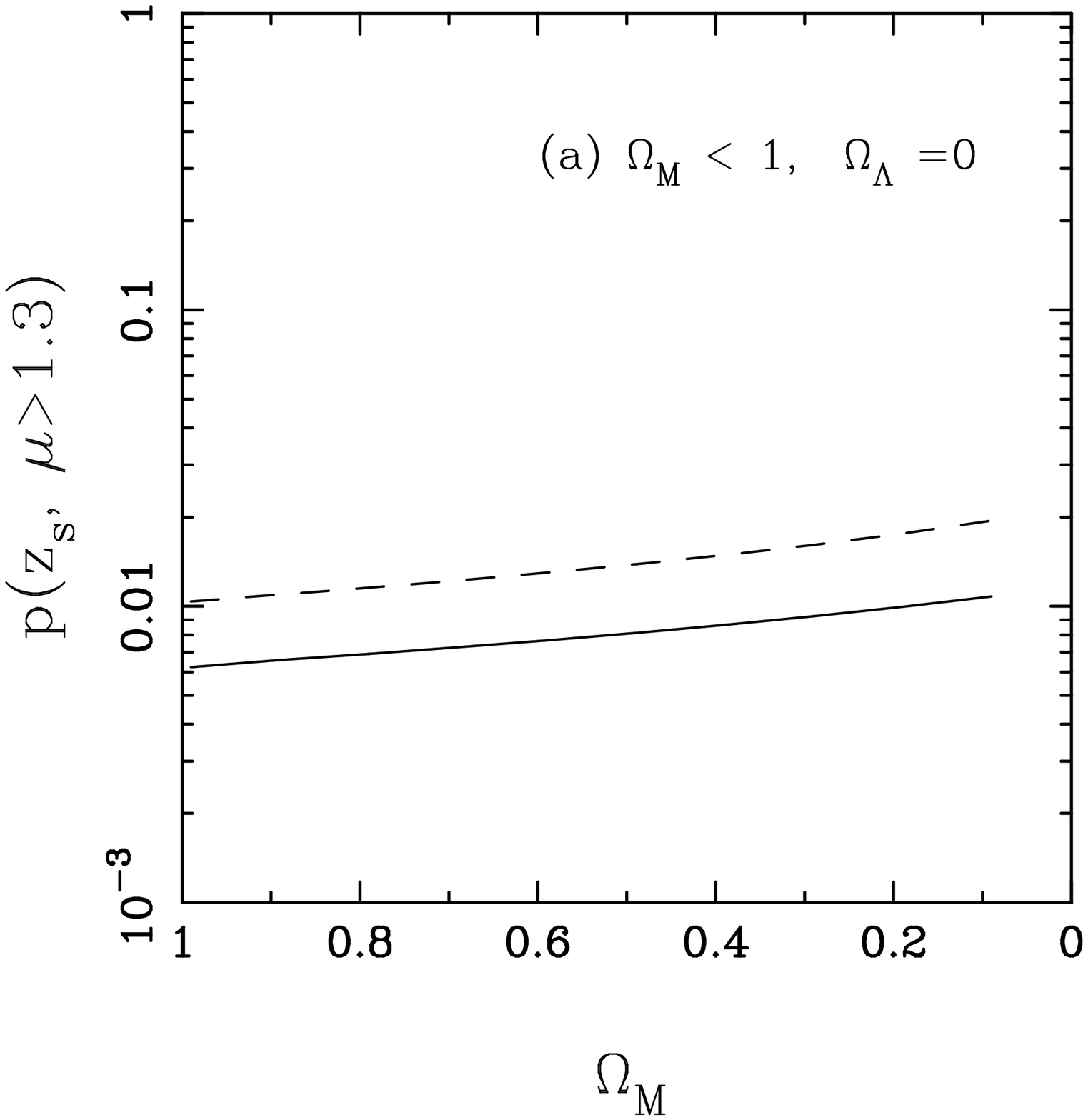,width=9.0cm}\hfill 
\psfig{figure=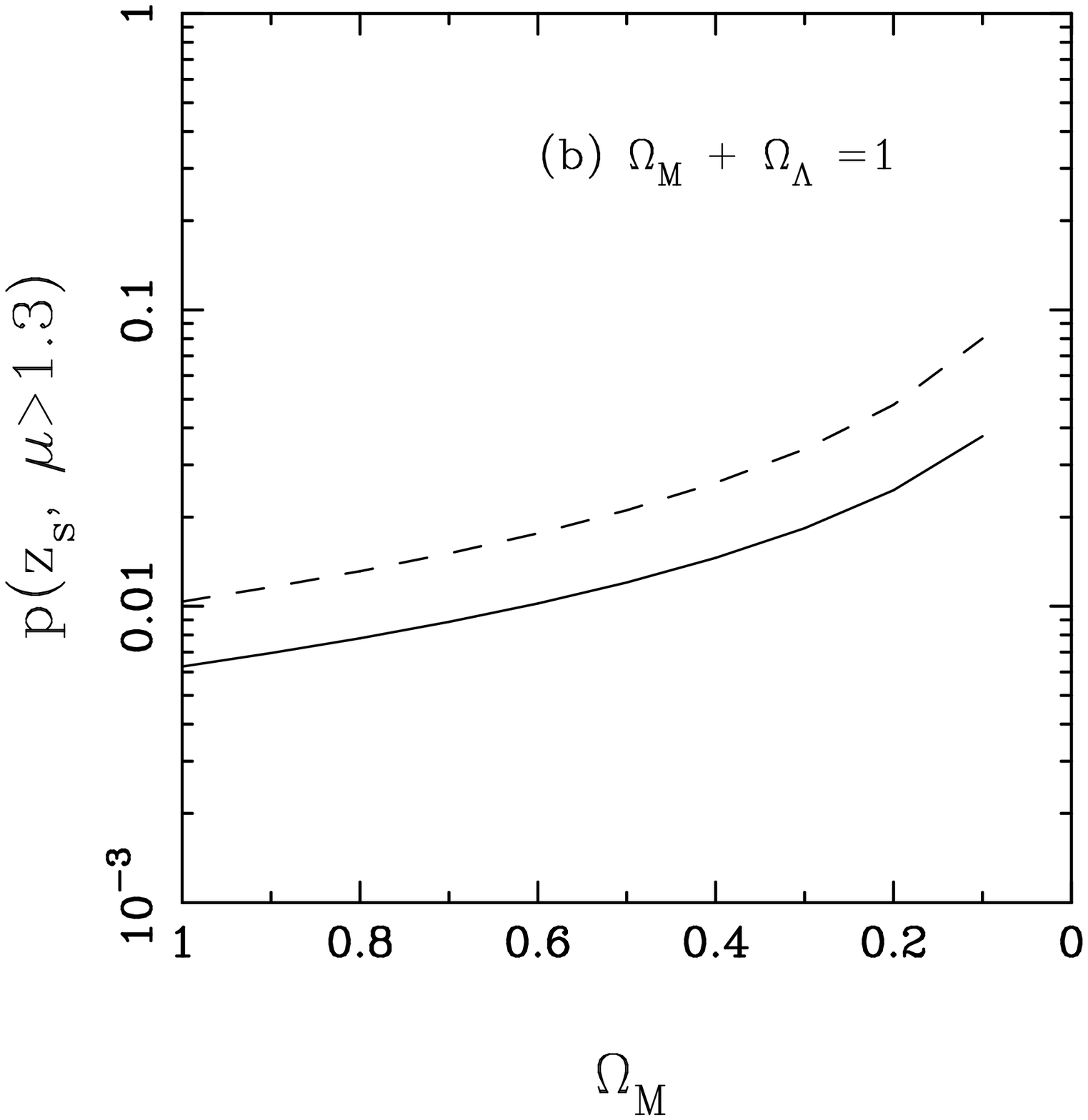,width=9.0cm}
}
        \caption{The probability, $p(z_s, \mu>1.3)$, that the sources located
		at $z_s=2$ (solid-lines) and $z_s=3$ (dashed-lines) 
		are magnified by the factor greater than $\mu=1.3$
		as a function of $\Omega_M$ for
	(a) an open universe with $\Omega_M < 1$, $\Omega_{\Lambda} =0$, and
	(b) a flat universe with $\Omega_M + \Omega_{\Lambda} =1$.  }
\end{figure*}

\begin{figure*}
\label{Figure 2}
\hbox{
\hspace*{0.6cm}
\psfig{figure=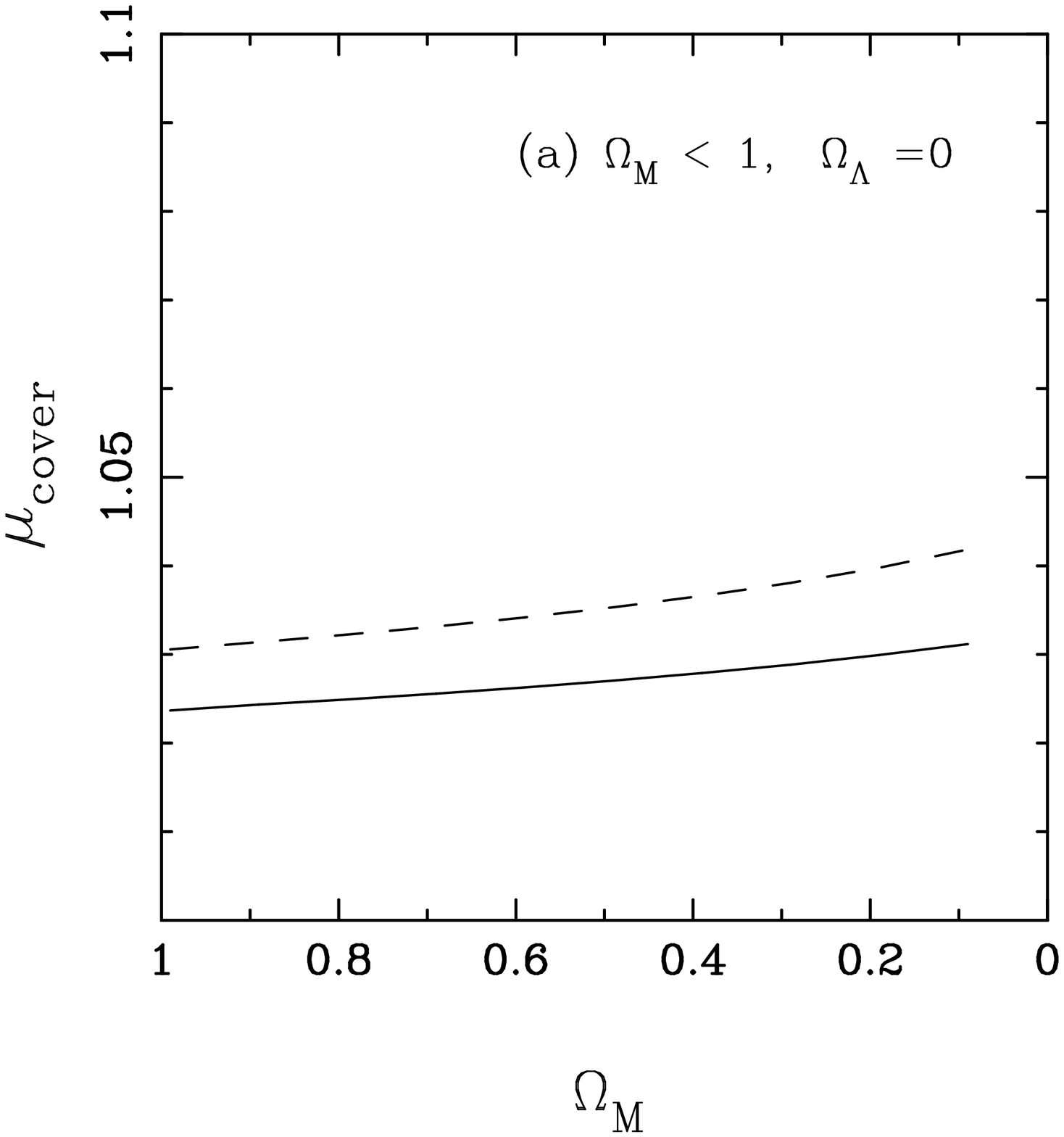,width=9.0cm}\hfill
\psfig{figure=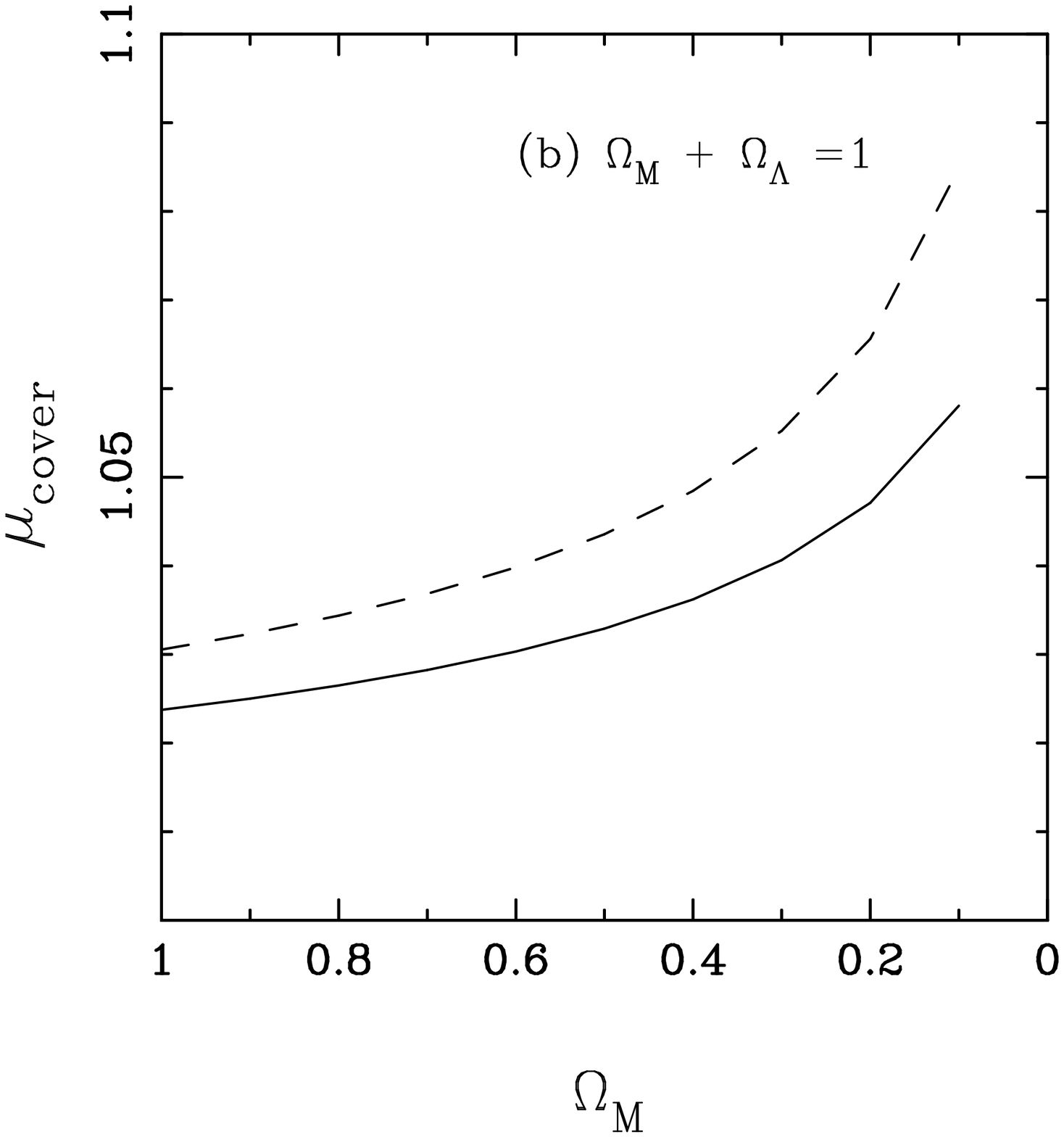,width=9.0cm}
}
        \caption{The magnification $\mu_{cover}$, which makes the probability 
		of the sources located at $z_s=2$ (solid-lines) and $z_s=3$ 
                (dashed-lines) $p(z_s, >\mu) = 1$,
                as a function of $\Omega_M$ for
        (a) an open universe with $\Omega_M < 1$, $\Omega_{\Lambda} =0$, and
        (b) a flat universe with $\Omega_M + \Omega_{\Lambda} =1$.  }
\end{figure*}

Fig.~1  shows how the probability $p(z_s, \mu>1.3)$ depends on the
normalized cosmological parameters $\Omega_M$ 
and  $\Omega_{\Lambda}$ for an open 
($\Omega_M < 1$, $\Omega_{\Lambda} =0$) 
and a flat ($\Omega_M + \Omega_{\Lambda}=1$)
universe respectively.
In our calculations, the source has been set at $z_s=2$ or $z_s=3$
respectively, and a moderate magnification of $\mu=1.3$ has been used.
Indeed, the probability $p(z_s, >\mu)$ depends sensitively 
on cosmological parameters $\Omega_M$ and $\Omega_{\Lambda}$.
However, even for a $\Lambda$-dominated ($\Omega_{\Lambda}=0.9$) flat universe,
only a small fraction ($<10\%$) of the sky can be moderately ($\mu \sim 1.3$) 
lensed by galaxies. 

Of course, by taking somewhat lower value for $\mu$ the probability of
the magnification can siginificantly increase. 
In order to get a more robust conclusion, we now estimate what value of 
magnification would affect the current observations of quasar count,
i.e., we calculate the value of $\mu_{cover}$ which makes $p(z_s, >\mu) = 1$.
The resulting magnification, which depends on both the cosmological model
and the source redshift, is shown in Fig.~2.
Since the magnification is generally much lower than $1.1$, Our result 
reinforces the hypothesis that the quasar counts are not seriously contaminated 
by the galactic lenses.

\begin{acknowledgements}
I thank X. P. Wu for helpful discussions and an anonymous refree for valuable
comments and suggestions.
This work was supported by the National Natural Science Foundation of China.
\end{acknowledgements}

\end{document}